\documentclass[doublecol]{epl2} 
\def\be{\begin{equation}}
\def\ee{\end{equation}}

\def\bc{\begin{center}} 
\def\ec{\end{center}}
\def\bea{\begin{eqnarray}}
\def\eea{\end{eqnarray}}
\newcommand{\avg}[1]{\langle{#1}\rangle}

\title{Superconductor-Insulator transition in a network of 2d percolation clusters}
\shorttitle{Superconductor-Insulator transition in a network of 2d percolation clusters} %Insert here a short version of the title if it exceeds 70 characters

\author{Ginestra Bianconi}
\shortauthor{Ginestra Bianconi}

\institute{School of Mathematical Sciences, Queen Mary University of London, London E1 4NS, United Kingdom}


\abstract{
 In this paper we characterize the superconductor-insulator phase transition on a network of 2d percolation clusters. Sufficiently close to the percolation threshold, for  $p\simeq p_c$, this network has a broad degree distribution,  and at $p=p_c$ the degree distribution becomes scale-free. 
We study the Transverse Ising Model on this complex topology in order to characterize the superconductor-insulator transition in a network formed by 2d percolation clusters of a superconductor material. We show, by a mean-field treatment, that the critical temperature of superconductivity depends on the maximal eigenvalue $\Lambda$ of the adjacency matrix of the network.
At the percolation threshold, $p=p_c$, we find that the maximal eigenvalue $\Lambda$ of the adjacency matrix of the network of 2d percolation clusters has a maximum. In correspondence of this maximum  the superconducting critical temperature $T_c$ is enhanced. These results suggest the design of new superconducting granular materials with enhanced critical temperature.}
\begin{document}

\maketitle

\section{Introduction}

Complex topologies strongly affect the phase diagram of classical phase transitions \cite{crit, Dyn}.
Scale-free networks, with power-law degree distribution $P(k)\sim k^{-\lambda}$ and diverging second moment of the degree distribution,  i.e. $\lambda\in(2,3]$, have a phase diagram  of the Ising model, or of the percolation phase transition,  which changes significantly with respect to the phase diagram of the same models on Poisson networks \cite{crit,Dyn}. Moreover,  the spectral properties of the networks drive  a number of other critical phenomena \cite{Durrett,epidemics2,Synchr1,Synchr2,Burioni_ON1,Burioni_ON2,Bradde}. 

Recently, quantum phase transitions defined on complex networks are starting to gain a growing attention.
Quantum critical phenomena depend on the topology of the underlying lattice, as it has been shown for Bose-Einstein condensation in heterogeneous networks \cite{Burioni_BEC}, for Anderson localization on scale-free networks with increasing cluster coefficient  \cite{Havlin_localization_1,Havlin_localization_2}, and for the Bose-Hubbard model on complex scale-free networks \cite{BH}.
Several papers have also  characterized  quantum processes on Apollonian networks \cite{JSAndrade2005,RFSAndrade2005}, which provide an example of scale-free networks embedded in two dimensions \cite{Hubbard_Apollonian,Free_Electron_Gas_Apollonian,Bose_Einstein_Apollonian}. 
In this context, the Transverse Ising Model is attracting increasing attention.

The random version of this model, the  Random Transverse Ising model, has been proposed to study the superconductor-insulator phase transition in granular materials \cite{IoffeMezard1,IoffeMezard2,QTIM}. In each grain of granular materials the superconducting order parameter is well defined, and the grains are coupled to each other by the  pair transfer term and the disorder is modulated by different on-site energies. Therefore,  the physics is similar to the superconductivity in Josephson junction arrays \cite{Fazio}.   In \cite{IoffeMezard1,IoffeMezard2} this model has been studied on a quenched Cayley tree network  by using  the quantum cavity method.
Subsequently, this model  has been studied on annealed complex networks \cite{QTIM},   i.e. networks that dynamically rewire their links. It  has been shown that the phase diagram  is strongly affected by a scale-free network topology of the underlying networks on which the model  is defined. In particular when  the second moment of the degree distribution $\avg{k^2}$ diverges with the network size, the critical temperature for the superconductor-insulator of this model diverges. This suggests that, by modulating the topology of the underlying network, the critical superconducting temperature can be enhanced. 

Nevertheless, the physical realization of complex networks as the underlying structure where condensed matter processes take place, poses many questions.
Here, we propose to model a 2d percolation pattern of superconducting materials as a network formed by interacting superconducting clusters.
This model is inspired by recent experimental data on the structure of
high temperature superconductors (HTS). Cuprates, diborides, and iron based
compounds are made of 2d superconducting layers, intercalated by spacer
layers, with a lattice misfit strain \cite{strain1,strain2} that induces
incommensurate phases \cite{Bak}. Here, defects self organization \cite{Littlewood,geballe} plays a
key role in the formation of complex nanoscale electronic heterogeneity
\cite{Zaanen}. A percolating network of 2d
superconducting grains has been observed in doped diborides \cite{Sharma},
in the electron doped iron chalcogenides \cite{Ricci1,Ricci2}, in
cuprates using scanning nano x-ray diffraction \cite{fratini, poccia2012},
and time resolved x-ray diffraction \cite{Poccia,demello2}.

These data show that  high temperature superconductivity (HTS) is also controlled
 by the spatial nanoscale arrangement of superconducting grains.
Therefore the theoretical focus for the mechanism of high $T_c$ is shifting
toward the search for the mechanism enhancing $T_c$ in a network of
superconducting grains in a 2d lattice \cite{Muller2,Kresin,Kresin2}.

In this scenario, the pseudogap phase, appearing in the
underdoped phase of cuprates at temperatures $T^{\star}$ above the superconducting
phase, is assigned to disconnected superconducting clusters formed at $T^{\star}$, which become
phase coherent at $T_c$ \cite{demello1}. The in plane resistivity of the
normal state (well below $T_c$ in the presence of very intense magnetic
fields used to destroy the superconducting state) exhibits a metal to
insulator (M-I) crossover \cite{Boebinger1,Boebinger2} at the same doping as the maximum of $T_c$.
This transition from an insulating to a metallic normal state is assigned
to nanoscopic superconducting grains in the underdoped phase, percolating
where the critical temperature for superconductivity reaches its maximum.
Therefore,  we extend our recent theoretical discovery of the increasing
$T_c$ in scale free networks \cite{QTIM} to the case of networks of
superconducting grains in a 2d lattice in the proximity of the percolation  threshold to be compared with experimental realizations.
In particular, we construct a network of 2d percolation clusters which are in close proximity. We show that at the percolation transition the network between clusters becomes scale-free.
Moreover,we show, by a mean-field calculation, that the critical temperature of the Quantum Transverse Ising Model, proposed to study the supercondutor-insulator transition in this system, has a maximum at the percolation threshold $p=p_c$ for any given size of the 2d dimensional array.
This provides a new roadmap for the design of complex superconducting materials with enhanced critical temperature $T_c$.

\section{Network of 2d percolation clusters}
\label{2d}

In order to mimic the fractal background present in cuprates \cite{fratini,poccia2012}, we consider a granular $2d$ material formed by percolating clusters of superconductors that could be realized in artificial arrays of superconducting grains connoted by Josephson Junctions on a 2D surface.
We consider a square lattice of size $L$ where each site is occupied by a grain of superconducting material with probability $p$. The array of grains displays a site percolation pattern \cite{Stauffer}.  Two occupied sites, $m$ and $n$ at distance $d_{mn}\leq \sqrt{2}$, belong to the same percolating cluster. In other words a cluster is formed by occupied nearest neighbors and next nearest neighbors. The percolation threshold for this case was studied in \cite{Galam}, where the percolation threshold was found to be $p_c=0.407\ldots$.

\begin{figure}
\begin{center}
\includegraphics[width=.5\columnwidth]{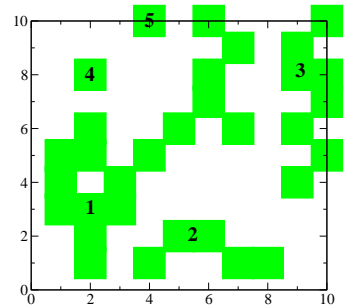}
\end{center}
\caption{(Color online)  Percolating clusters in an array of size $L=10$ with $p=0.30$. The number indicates the labelling of each to the five clusters. In a coarse-grained picture any two percolating clusters  $i$ and $j$ at distance $d_{ij}\leq \sqrt{5}$ are connected together. Therefore cluster $1$ is connected to all the other clusters and no other links are present in this network. The adjacency matrix of the network is given by Eq. $(\ref{adj})$.}
\label{per_10}
\end{figure}

\begin{figure}
\begin{center}
\includegraphics[width=.95\columnwidth ]{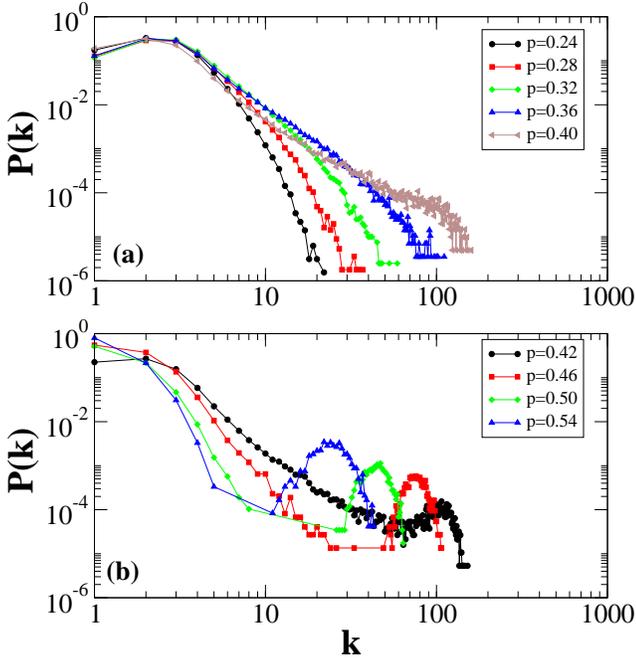}
\end{center}
\caption{(Color online)   The maximal eigenvalue of the network as a function of $p$ for different linear sizes $L$  of the array. The data shown are generated from 2d array of linear size $L=100$ and the data are averaged over $1000$ realizations of these networks. The  transition for this array of linear size $L=100$ is about ${p}_c=0.407\ldots$.}
\label{denk}
\end{figure}
Here, we define the distance between two clusters, $i$ and $j$, as the minimal distance between any site $n$ belonging to  the first cluster and any site $m$  belonging to the second cluster, i.e.
\begin{equation}
d_{i,j}=min_{n\in {\cal C}(i),m\in {\cal C}(j)}d_{nm}
\end{equation}
where ${\cal C}(i)$ are the set of sites belonging to the cluster $i$.
In order to have a coarse grained view of this system we construct a network of percolating clusters.
Each percolating cluster $i$ is linked to any other percolating cluster $j$ which is in close proximity.
There are different ways to implement this definition. Here we consider two percolating clusters, $i$ and $j$ linked, if their distance satisfy the following requirements:
\begin{equation}
2\leq d_{ij}\leq \sqrt{5}.
\label{cond}
\end{equation}
In Figure $\ref{per_10}$ we show a small array of size $L=10$ and  the percolation clusters obtained in the method mentioned  for $p=0.30$. In this figure we have labelled the five percolation clusters with an integer number  from 1 to 5. Each pair of percolating clusters $i$ and $j$ are linked in a network if the distance between the cluster satisfies the condition given in Eq. $(\ref{cond})$.
The adjacency matrix $a$ of the coarse-grained percolation network, that models the interaction of the clusters in Figure $\ref{per_10}$, is given by
\bea
a=\left(\begin{array}{c cc c c}  0& 1& 1&1 &1\\
1&0 & 0& 0 &0 \\
1&0 & 0& 0 &0 \\
1&0 & 0& 0 &0 \\
1&0 & 0& 0 &0 \\
\end{array}\right).
\label{adj}
\eea

\begin{figure}
\begin{center}
$\begin{array}{c}
\includegraphics[width=.70\columnwidth]{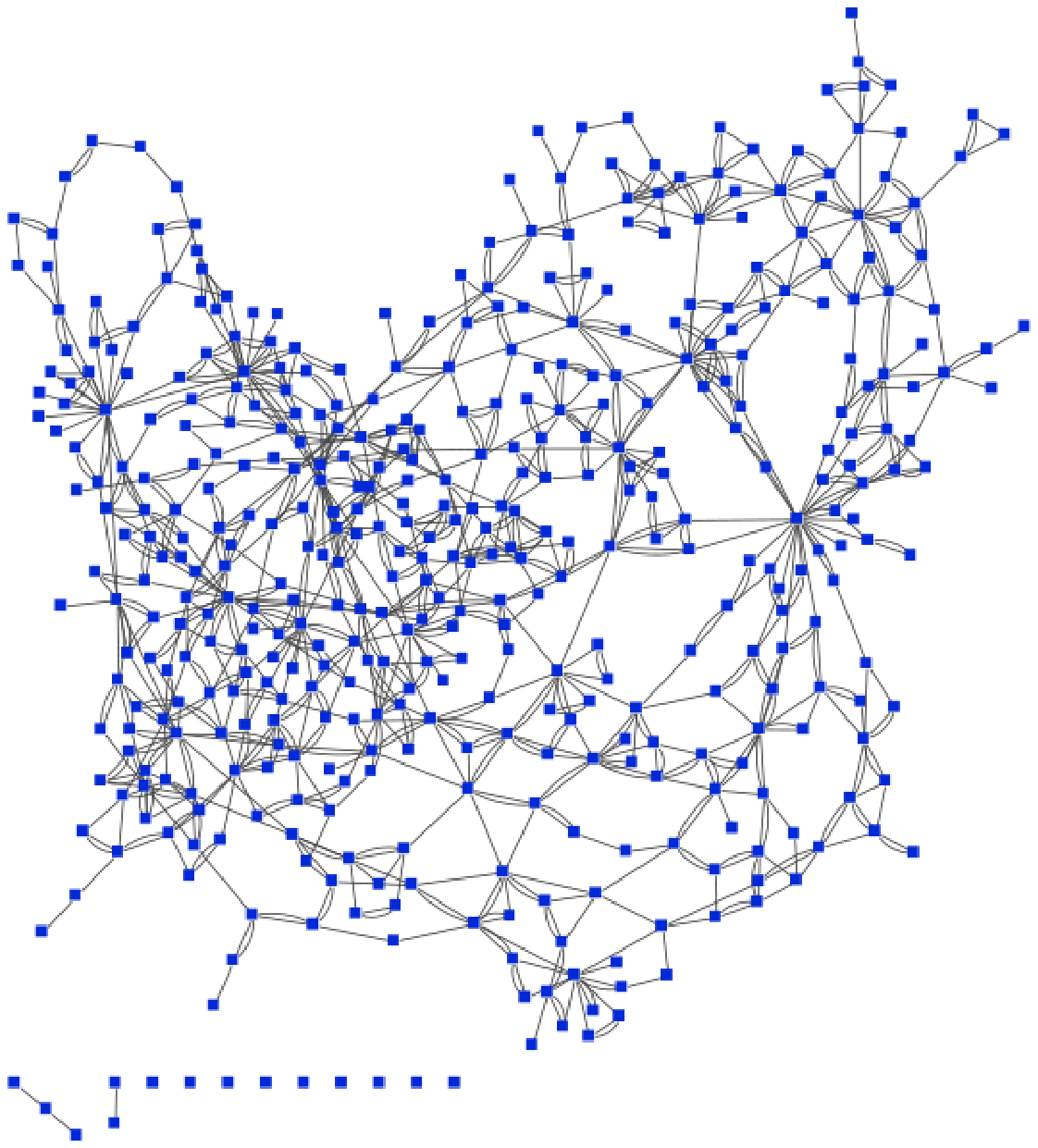}\nonumber \\
\includegraphics[width=.70\columnwidth]{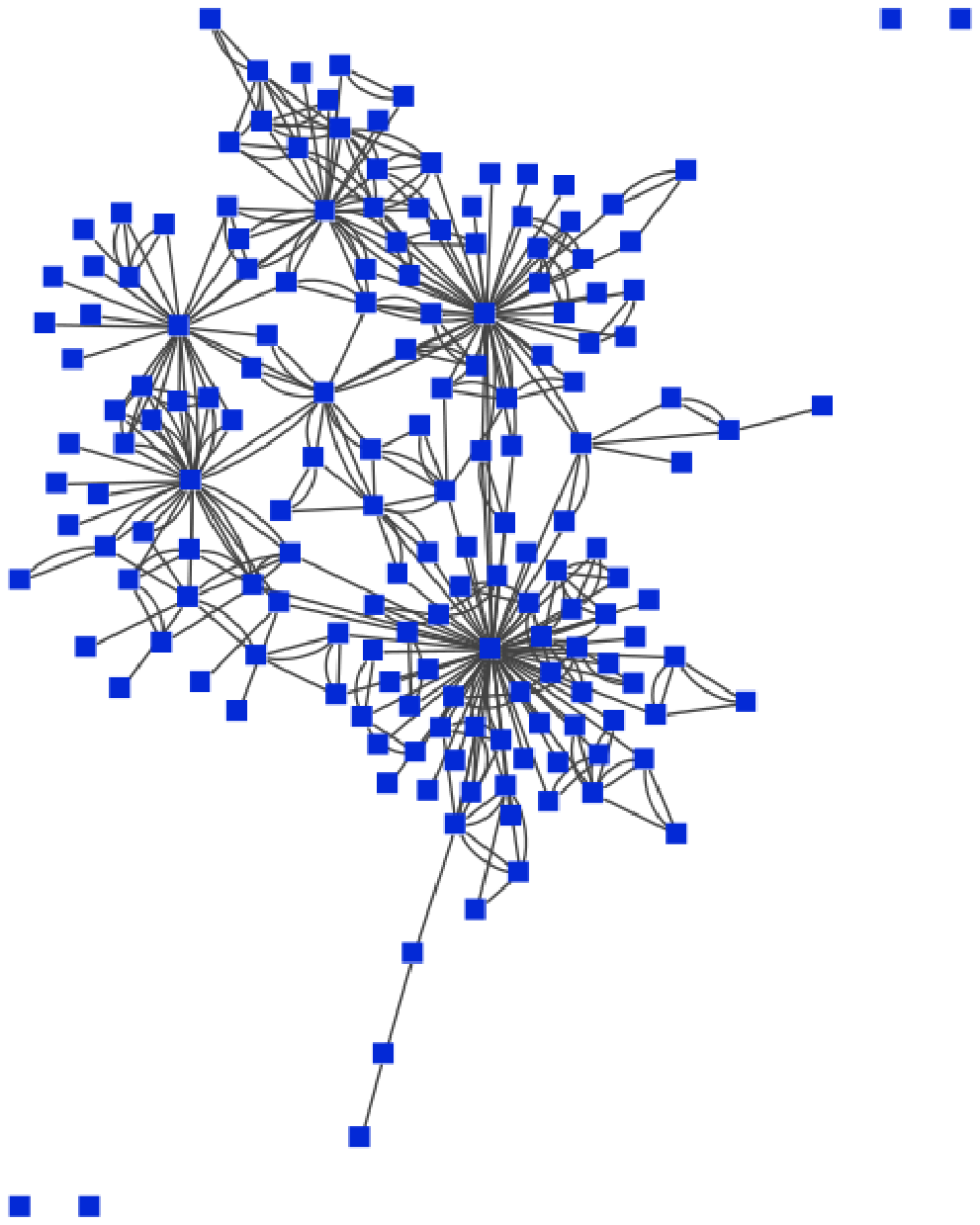}\nonumber \\
\includegraphics[width=.75\columnwidth]{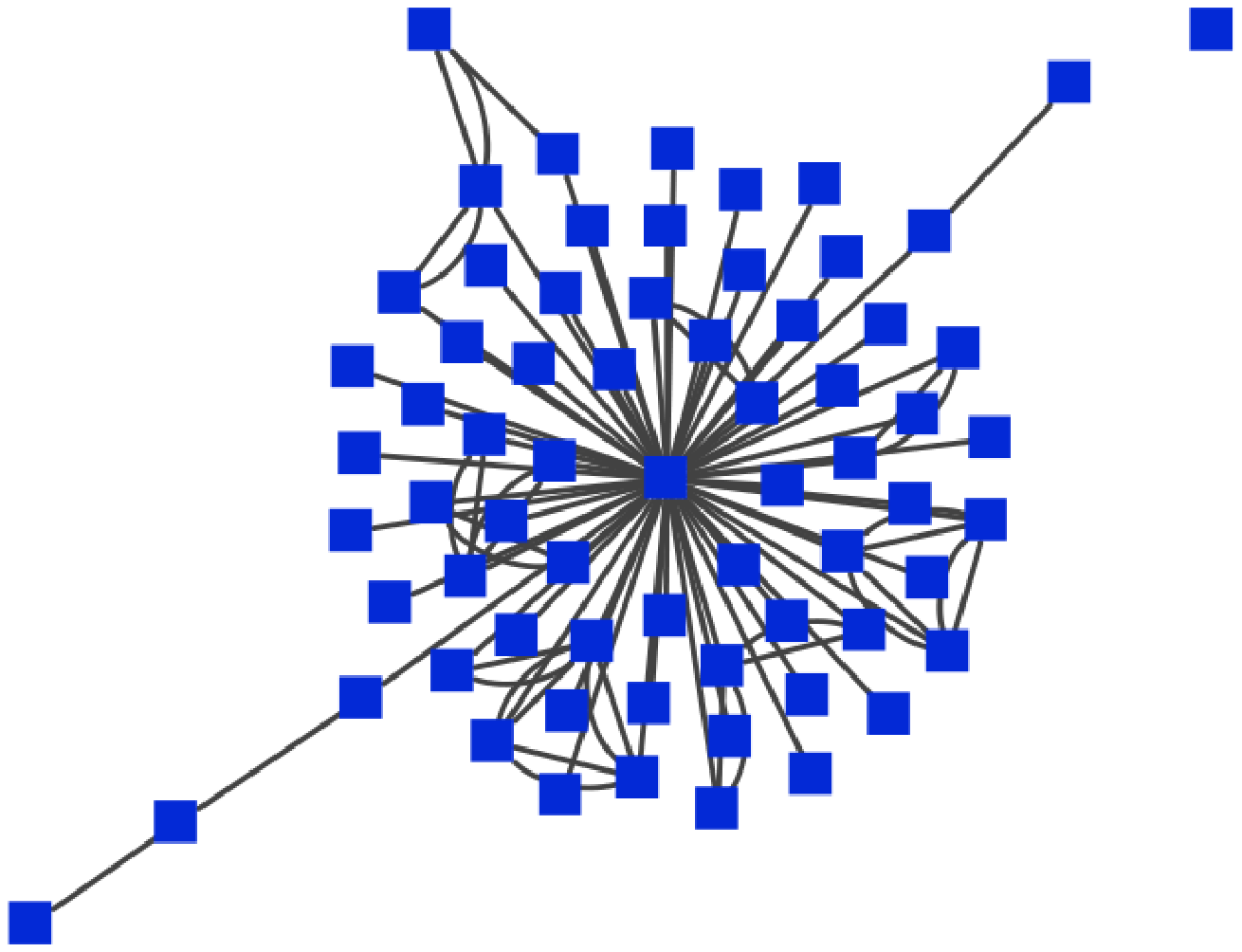}\nonumber \\
\end{array}$
\end{center}
\caption{(Color online)  The network between 2d percolating clusters. From the top to the bottom we show typical networks  for $p=0.32,0.40,0.48$.}
\label{net}
\end{figure}

At the percolation transition it is well known that the percolating clusters becomes fractal \cite{Stauffer}.
Moreover it is well known that the distribution $P(s)$ of the size $s$ of the percolating clusters  will scale like 
\bea
P(s)\sim s^{-\tau}\Phi(s/s_c)
\eea
with $s_c=|p-p_c|^{-\sigma}$ \cite{Stauffer}.
This distribution becomes a pure power-law distribution at $p=p_c$.
As we change the parameter $p$, we can study the degree distribution of coarse grained network.
What we observe is that, as we approach the percolation transition, the degree distribution of the coarse grained network becomes scale-free.
This is intuitively explained by the fact that larger clusters will have a larger perimeter, and therefore more possibilities to link with nearby percolating clusters.
In Figure $\ref{denk}$  we show the degree distribution of the network for $p<p_c$ and $p>p_c$ respectively. For $p<p_c$ the degree distribution displays an exponential cutoff. For $p>p_c$ the degree distribution is dominated by hub clusters with a very high degree present in the network. Finally for $p=p_c$ the degree distribution is scale-free.
In Figure $\ref{net}$ we show the network of 2d percolating clusters as a function of $p$ for a value of $L=100$. For $p<p_c$ the network has many clusters with relatively low connectivity, at $p\simeq p_c$ a hierarchy of hub nodes appear, finally for $p>p_c$ the number of cluster is strongly reduced and one hub node dominates the structure of the network.

\begin{figure}
\begin{center}
\includegraphics[width=.6\columnwidth]{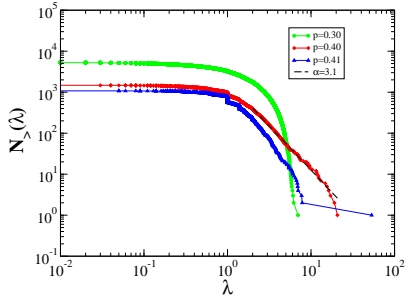}
\end{center}
\caption{(Color online)  
The  spectrum of a single network formed by a 2d percolation cluster of and array of linear size $L=500$. In particular we plot the rank $N_{>}(\lambda)$ of the eigenvalue $\lambda$ against $\lambda$. 
The rank of the eigenvalues is proportional to the cumulative distribution  $N_{>}(\lambda)\propto P_{>}(\lambda)$ of the eigenvalues $\lambda$. We show that, at the percolation threshold, when $p\simeq p_c$, the rank of the eigenvalues scale as a power-law, i.e.  $N_>(\lambda)\propto \lambda^{-\alpha+1}$.  This implies that the probability density of the eigenvalue goes like $\rho(\lambda)\propto \lambda^{-\alpha}$ with $\alpha\simeq=3.1$
}
\label{spectrum}
\end{figure}

\section{Mean field calculation  of the  Transverse Ising model the  quenched  networks of 2d percolating clusters}

In this section we propose to study the Transverse Ising model to characterize the superconductor-insulator phase transition in granular superconductors close to the percolation threshold.
We will adopt a coarse grained view of the 2d percolative clusters in which each cluster of superconductor material is a node of a network, and there is a coupling between nearby percolating clusters.
The fact that the degree distribution of the network of 2d percolating clusters is scale-free at the percolation transition is suggestive, since it was shown that on annealed scale-free network the critical temperature of the superconductor-insulator phase transition is diverging with the network size \cite{QTIM}.
Nevertheless, the network under consideration here is both quenched and non-random, and we need to investigate further the topology of the model in order to make some conclusions on the critical temperature of the superconductor-insulator phase transition defined on this network.

We consider a system of  spin variables 
$\sigma_i^x,\sigma_i^z$, for $i=1,\dots,N$, defined on the nodes of a given quenched network  (the 2d percolating clusters) with adjacency matrix ${\bf a}$.
The Traverse Ising Model is defined as
 \be 
 \hat{H}=-\frac{J}{2}\sum_{ij}a_{ij}\hat{\sigma}^x_i \hat{\sigma}^x_j-\sum_i \epsilon \hat{\sigma}^z_i.
\label{H0}\ee
This Hamiltonian is a simplification respect to the $XY$ model Hamiltonian proposed by Ma and Lee \cite{Ma} to describe the superconducting-insulator phase transition but to the leading order the mean-field  equations for the order parameter are  the same, as widely discussed in \cite{IoffeMezard1, IoffeMezard2}.
The Hamiltonian describes the superconducting-insulator phase transition as a ferromagnetic spin $1/2$ spin system in a transverse field.
We propose to use this Hamiltonian to describe, in a granular superconductor, the transition from a phase of superconducting grains with no phase coherence (called insulator for granular superconductors)  to the low temperature superconducting phase with phase coherence. Each node of the network corresponds to a 2d percolation cluster and the adjacency matrix is the adjacency matrix of the network of 2d percolation clusters described in the previous section.
{ In this model we consider clusters  with a minimun size large enough so that there are no large charging effects 
 and the coherence lenght is always smaller than their size. For superconductors with short coherence lenght the minimum size could be about 10-20 nanometers.}
The spins  $\sigma_i$  in Eq. $(\ref{H0})$ indicate occupied or unoccupied states by Cooper pairs or localized pairs; the parameter $J$ indicates the couplings between neighboring spins, $\epsilon$  is the  on-site energy.   Finally, in  this model the superconducting phase corresponds to the existence of a  spontaneous magnetization in the $x$ direction.  
{ We could also consider a different version of the model in which the network of interaction is weighted, and  the link's weight between two nodes depends on the number of sites in the two clusters that are at distance $d_{ij}\in [2,\sqrt{5}]$. We have checked that this modified  version of the model doesn't change the main result of the paper. Therefore, for simplicity, in this paper we consider only simple networks with adjacency matrix elements $a_{ij}=0,1$.}
Close to the percolation threshold, the spectrum of the adjacency matrix of the network of 2d percolation clusters develops a power-law tail $\rho(\lambda)\propto \lambda^{-\alpha}$ (See figure \ref{spectrum}), 
therefore losing memory of the underlying 2d dimensional structure.
Therefore in  order to study the dynamics of the granular superconductor, we perform a mean-field approximation in which we put
\bea
\hat{\sigma}^x_i {\hat{\sigma}^x_j}\simeq \hat{\sigma}^x_i \avg{\hat{\sigma}^x_j}+\avg{\hat{\sigma}^x_i}{\hat{\sigma}^x_j}-\avg{\hat{\sigma}^x_i} \avg{\hat{\sigma}^x_j}.
\eea
We can therefore consider the following mean-field Hamiltonian, in which we use $m_i^x=\avg{\hat{\sigma}^x_i}$ 
 \be 
 \hat{H}_{MF}=-J \sum_{ij}a_{ij}\hat{\sigma}^x_i m_i^x+\frac{J}{2}\sum_{ij}a_{ij} m_i^x m_j^x-\sum_i \epsilon \hat{\sigma}^z_i.
\label{HMF}\ee
In the mean field approach the partition function for this problem is given by 
\be
Z=\mbox{Tr}\  e^{-\beta \hat{H}_{MF}}
\ee
with the Hamiltonian given by Eq. $(\ref{HMF})$.
The mean-field Hamiltonian can be written as
\bea
\hat{H}_{MF}=\hat{E}+\frac{J}{2}\sum_{ij} a_{ij} m_i^x m_j^x\eea
with 
\bea
\hat{E}=-J\sum_{i}\hat{\sigma}^x_i h^{MF}_i-h\sum_i \hat{\sigma}_i^{x},
\eea
where the local fields $h_i^{MF}$ are given by 
\begin{equation}
h_i^{MF}=\sum_j a_{ij} m_j^x.
\label{hiMF}
\ee
For this problem the magnetizations along the axis $x$,  $m_i^x$, and along the axis $z$,  $m_i^z$, can be calculated by evaluating 
\bea
m_{i}^x=\frac{\mbox{Tr}\sigma_{i}^x e^{-\beta \hat{H}^{MF}}}{Z}\nonumber \\
m_{i}^z=\frac{\mbox{Tr}\sigma_{i}^z e^{-\beta \hat{H}^{MF}}}{Z}.
\eea 
Performing these calculations we get
\bea 
\hspace*{-8mm}m_{i}^x&=&\frac{Jh_i^{MF}}{\sqrt{(Jh_i^{MF})^2+\epsilon^2}}\tanh\left(\beta \sqrt{(Jh_i^{MF})^2+\epsilon^2}\right),\label{Self} \\
\hspace*{-8mm}m_{i}^z&=&\frac{\epsilon}{\sqrt{(Jh_i^{MF})^2+\epsilon^2}}\tanh\left(\beta \sqrt{(Jh_i^{MF})^2+\epsilon^2}\right). \label{Mz}
\eea
The Eqs. $(\ref{Self})$ are the self-consistent equations that together with Eqs. $(\ref{hiMF})$ determine the local magnetizations $m_i^x$.
Close to the phase transition $m_i^x\ll1$ the self-consistent equation for the magnetization $m_i^x$ is given by
\bea 
m_{i}^x&=&{J\sum_j a_{ij} m_j^x}\frac{\tanh(\beta \epsilon)}{\epsilon}.
\eea
If we diagonalize this equation along the eigenvalues of the adjacency matrix $a$, we find that the phase transition occurs at 

\be
1=J\Lambda \frac{\tanh(\beta \epsilon)}{\epsilon}
\label{Tc}
\ee
where $\Lambda$ is the maximal eigenvalue of the adjacency matrix ${\bf a}$ of the network.

\begin{figure}
\begin{center}
\includegraphics[width=.6\columnwidth]{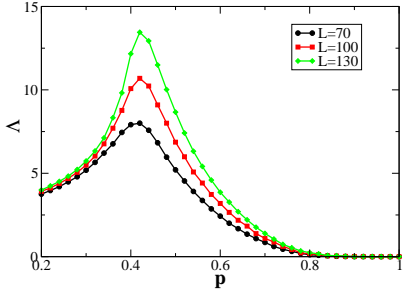}
\end{center}
\caption{(Color online) The maximal eigenvalue $\Lambda$ of the network as a function of $p$ for different linear sizes $L$  of the array.  Data are averaged over 500 realizations.}
\label{Lambdap}
\end{figure}

\begin{figure}
\begin{center}
\includegraphics[width=.6\columnwidth]{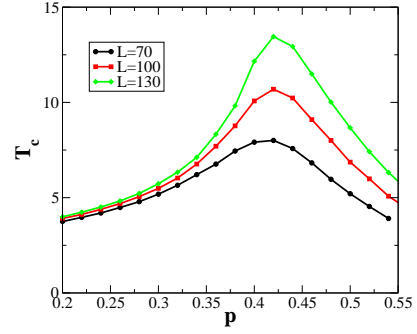}
\end{center}
\caption{(Color online)   Critical temperature $T_c$ of the superconductor-insulator phase transition given by Eq. $(\ref{Tc})$ with $J=1$ and $\epsilon=1$ plotted as a function of $p$. The data are averaged over 500 realizations.}
\label{Tc.fig}
\end{figure}

Therefore  for $\Lambda\to \infty$ with $N\to \infty$ then $\beta\to 0$  for any fixed value of the coupling $J>0$,  and the critical temperature for the paramagnetic ferromagnetic phase transition $T_c$ diverges.
\section{Superconductor-insulator phase transition on a network of 2d percolation clusters}

In order to study the superconducting insulator phase transition on granular materials close to percolation, we consider the network formed by 2d percolating clusters.
The critical temperature $T_c$ of the superconductor-insulator phase transition provided by the mean-field calculation Eq. $(\ref{Tc})$, depends on the maximal eigenvalue of the adjacency matrix $\Lambda$ of the network.
Therefore, we have studied numerically the maximal eigenvalue of the adjacency matrix of the network of 2d percolating clusters.
In Figure $\ref{Lambdap}$ we show the maximal eigenvalue $\Lambda $ as a function of $p$, and we observe that this eigenvalue has a maximum at $p=p_c=0.407\ldots$ for $L=100$.
Moreover, in this figure we show that in  the network of 2d percolating clusters, sufficiently close to ${p}_c$, the maximal eigenvalue of the matrix $\Lambda\to \infty$ as $N\to \infty$. Moreover, for every fixed value of $N$, we have that the maximal eigenvalue $\Lambda$ has a peak  for $p={p}_c$.
In Figure $\ref{Tc.fig}$ we show the value of the critical temperature given by Eq. $(\ref{Tc})$ for the superconductor-insulator phase transition on the network of 2d percolation clusters as a function of $p$ showing a peak for $p={p}_c$.
Therefore we predict that the system will display a maximum of the superconductor critical temperature at percolation.

\section{Conclusions}
In this paper we have investigated how a granular material close to percolation can be described at the coarse-grained level as a complex network. We have shown that this complex network can be constructed by linking together the   next-nearest clusters. Close to the percolation threshold, the distribution of the sizes of the cluster is power-law distributed and also the degree distribution of the coarse-grained network is scale-free.  Moreover, the adjacency matrix of the  2d percolation network has a maximal eigenvalue that diverges with the network size close to the percolation transition.
In this coarse-grained network picture of interaction between nearby  percolation clusters, we have studied the mean-field of a Transverse Ising model. This model is proposed in order to characterize the superconductor-insulator transition in a 2d array close to percolation.
We have shown that if the maximal eigenvalue of the adjacency matrix of the network diverges, the critical temperature of the system diverges as well.
Therefore with this coarse-grained network model we predict that granular array  show a maximum of the critical temperature for the superconductor-insulator phase transition at the percolation threshold.

The present theory can now be tested producing artificial geometrical arrays of 2d superconducting grains on a flat surface. In fact, artificial arrays made of mesoscopic superconducting islands placed on normal metal surface can now be made by using available advanced nanotechnology \cite{Eley}. These artificial arrays today provide practical tunable realizations of two-dimensional granular (2d) superconductivity. An alternative approach is the production of artificial superconducting puddles embedded in a 2d layer intercalated by different block layers in a layered material. The manipulation of the arrays of superconducting grains can be done by controlling defects self-organization by external fields \cite{Poccia,demello2}. Finally these results open new perspectives for materials scientists looking for new room temperature superconductors by focusing on materials design by pattering at the nanoscale.

%\acknowledgments
%Insert here the text.

\end{document}